\documentclass[useAMS,usenatbib,usegraphicx]{mn2e}

\def \aj {AJ}
\def \mnras {MNRAS}
\def \apj {ApJ}
\def \apjs {ApJS}
\def \apjl {ApJL}
\def \aap {A\&A}
\def \aaps {A\&AS}

\def \pasp {PASP}
\def \aaps {AAPS}
\def \angstrom {{\rm \AA}}

\usepackage{amssymb}
\usepackage{lscape}

\title[M32 composite stellar population]
 {An analysis of the composite stellar population in M32\thanks{
Based on observations obtained at the Gemini Observatory, which is operated by the Association of Universities for Research in Astronomy, Inc., under a cooperative agreement with the NSF on behalf of the Gemini partnership: the National Science Foundation (United States), the Particle Physics and Astronomy Research Council (United Kingdom), the National Research Council (Canada), CONICYT (Chile), the Australian Research Council (Australia), CNPq (Brazil), and CONICET (Argentina) 
(observing run ID: GN-2004B-Q-74).}}
\author[P. Coelho, C. Mendes de Oliveira and R. Cid Fernandes]
{P.~Coelho$^{1}$,
C.~Mendes de Oliveira$^2$,
R.~Cid Fernandes$^3$\\
$^1$ Institut d'Astrophysique, CNRS, Universit\'e Pierre et Marie Curie, 98 bis Bd Arago, 75014 Paris, France (pcoelho@iap.fr)\\
$^2$ Universidade de S\~ao Paulo, IAG, Rua do Mat\~ao 1226,
 S\~ao Paulo 05508-900, Brazil (oliveira@astro.iag.usp.br)\\
$^3$ Departamento de F{\'i}sica, CFM, Universidade Federal de Santa Catarina, PO Box 476, 88040-900 Florian{\'o}polis, SC, Brazil (cid@astro.ufsc.br)\\}
\date{}

\pagerange{\pageref{firstpage}--\pageref{lastpage}} \pubyear{2006}

\def\LaTeX{L\kern-.36em\raise.3ex\hbox{a}\kern-.15em
 T\kern-.1667em\lower.7ex\hbox{E}\kern-.125emX}

\begin{document}

\label{firstpage}

\maketitle

\begin{abstract}
We obtained long-slit spectra of high signal-to-noise ratio of the galaxy M32 with the 
Gemini Multi-Object Spectrograph at the Gemini-North telescope. 
We analysed the integrated spectra by means of full spectral fitting
in order to extract the mixture of stellar populations that best represents its composite nature.
Three different galactic radii were analysed, from the nuclear region out to 2 arcmin from the centre. 
This allows us to compare, for the first time, the results of integrated light spectroscopy 
with those of resolved colour-magnitude diagrams from the literature.
As our main result, we propose that an ancient and an intermediate-age population co-exist in M32,
and that the balance between these two
populations change between the nucleus and outside one effective radius (1\,$r_{\rm eff}$) in the sense that
the contribution from the intermediate population is larger at the nuclear region. 
We retrieve a smaller signal of a young population at all radii
whose origin is unclear and may be a 
contamination from horizontal branch stars, such as the ones identified by Brown et al. 
in the nuclear region.
We compare our metallicity distribution function for a region 1 to 2 arcmin
from the centre to the one obtained with photometric data by Grillmair et al. Both
distributions are broad, but our spectroscopically derived distribution has a significant
component with $[Z/Z_{\sun}] \leq -1$, which is not found by Grillmair et al.

\end{abstract}


\section{Introduction}

\begin{table*}
\caption{Parameter coverage of the fits. Adopted: Z$_{\sun}$ = 0.017.}
\label{tab_models}
\begin{center}
\begin{tabular}{lclcc}
\hline
Model & log(age[yr]) & \multicolumn{1}{|c|}{[Z/Z$_{\sun}$]} & Wavelength range (\angstrom)\\
\hline
BC03               & 8.0 $-$ 10.3 & $-$2.2, $-$1.6, $-$0.6, $-$0.3, $+$0.05, $+$0.5 &  4700 $-$ 8750\\ 
Charlot \& Bruzual & 8.0 $-$ 10.3 & $-$2.2, $-$1.6, $-$0.6, $-$0.3, $+$0.05, $+$0.5 &  4700 $-$ 7400\\ 
PEGASE-HR          & 8.0 $-$ 10.3 & $-$1.6, $-$0.6, $-$0.3, $+$0.1, $+$0.05 &        4700 $-$ 6800\\ 
Vazdekis et al.    & 8.0 $-$ 10.3 & $-$1.6, $-$1.2, $-$0.6, $-$0.3, $+$0.05, $+$0.25 & 4700 $-$ 7400\\ 
\hline
\end{tabular}
\end{center}
\end{table*}

M32 is a controversial galaxy. This low-mass satellite of M31 has long been considered the prototype of the {\it compact elliptical} morphological classification \citep{bender+92}: low mass, high surface brightness galaxies, tidally truncated companions to massive galaxies. So far, few objects match this description \citep{ziegler_bender98,mieske+05,chilingarian+07}. The origins of the peculiar structural properties of M32 are still a matter of debate and the proposed models span a wide range of hypotheses: from a true
elliptical galaxy at the lower extreme of the mass sequence to a threshed early-type spiral
\citep{nieto_prugniel87,bekki+01,choi+02,graham02,gallaguer+05,young+08}. 

Controversial or not, M32 is a galaxy we need to understand
because it plays an important role in resolving controversies about how 
to interpret the integrated spectra of galaxies: it is an early-type 
galaxy where we can get extremely high signal-to-noise (S/N) in the spectra, good coverage outside 
the nucleus to look for radial population gradients and resolve individual stars to 
a very meaningful magnitude level. If we hope to claim that we understand the evolution of distant early-type galaxies, 
we must first be able to demonstrate an understanding of M32 stellar population.

The stellar population of M32 has been extensively studied by means of
integrated spectroscopy and resolved photometry. Most studies of the integrated spectrum of M32 have found 
signatures of an intermediate-age and close to solar metallicity population
\citep[e.g.][]{rose85,davidge90,freedman92,jones_worthey95,vazdekis_arimoto99,delburgo+01,schiavon+04,worthey04,rose+05}. 
The most commonly used technique of deriving mean-ages and mean-metallicities from integrated spectroscopy, 
as articulated by \citet{worthey94}, consists in comparing selected spectral indices to the predictions
of simple stellar populations (SSP) models. 
Quoting an SSP-age and an SSP-metallicity, however, does not necessarily imply that a galaxy 
{\it is} an SSP. In fact, \citet{bica+90} and \citet{hardy+94} have shown that the spectral energy distribution
of M32 cannot be adequately fit by SSP models, and \citet{bica+90} argued that
oversimplifications might explain some of the conflicting results found in
literature 
\citep[see also][]
{oconnell80,schmidt+89,magris_bruzual93,rose94,trager+00b}.

Resolved photometric studies include those of \citet{alonsogarcia+04} 
and \citet{brown+00} who found signatures of a true ancient population in M32 [by the identification of RR-Lyrae
and an extended horizontal branch (HB), respectively], and \citet{davidge_jensen07} and references therein, who identified signatures of an intermediate age population. 
The most comprehensive study to date of the resolved stellar population of M32 has been carried out by \citet{grillmair+96}, who resolved individual stars down to slightly below the level of the HB with the 
\textit{Hubble Space Telescope} (HST) \textit{Wide-Field Planetary Camera 2} (WFPC2) in a region 1$-$2 arcmin from the centre of the galaxy. The most striking
feature of this colour-magnitude diagram \citep[CMD; and also the one by][ at a larger radii]{worthey+04}
is its composite nature, implying a range in [Z/H] from roughly solar down to below $-$1.0 dex. 

Spreads in age and Z are as important as the mean values in understanding
the star formation history of a galaxy, and a study that takes into account 
the composite stellar population of M32 may lead to a better understanding
of its origin and evolution. 
In order to address this question,
we obtained spectra of M32 along its major axis out to a radius of 2 arcmin, with the Gemini-North Telescope and Gemini Multi-Object Spectrograph (GMOS).
We adopt the code {\sc starlight}
\citep{cid+05} and state-of-the-art stellar population spectral models 
(\citealt{BC03}, hereafter BC03; \citealt{PEGASE-HR}, hereafter PEGASE-HR; 
Charlot \& Bruzual in preparation, hereafter Charlot \& Bruzual; 
Vazdekis et al. in preparation, hereafter Vazdekis et al.)
to perform a pixel-by-pixel modelling of the integrated spectrum of M32. 
For the first time we extend
the analysis of integrated spectra of M32 to the radii of the resolved 
CMD study by \citet{grillmair+96}.

Our observational data are described in Section 2, and we explain the analysis method in Section 3. 
Results are given in Section 4 and conclusions in Section 5. 

\section{Observational data}

Long-slit spectra of M32 were obtained on 2004 October 10, 11 and 16
with the GMOS on Gemini-North under the Program GN-2004B-Q-74. 
We used the R400 grating, a 0.75-arcsec slitwidth and an effective target 
wavelength at 6800\,\angstrom, resulting 
in spectra with coverage 4700 $-$ 8930\,\angstrom~and an average resolution of
full width at half-maximum (FWHM) = 5.4\,\angstrom. 

Four exposures of 30\,s each were taken with the slit positioned over the nucleus of the galaxy, 
along the major axis of the galaxy [position angle (PA) $=$ 165$^\circ$].
The slit was subsequently moved along the major axis off the central region to avoid contamination by scattered light
\citep[see][]{rose+05}, and four long exposures of 3060\,s each were taken. 
The basic reduction tasks $-$ trimming, bias and flat
field correction $-$ were done with the Gemini reduction package available for {\sc iraf}. 
The extraction of the spectra was then performed using the task {\sc apall} under the 
National Optical Astronomy Observatories (NOAO)
long-slit reduction package. 
The spectra were extracted using the variance-weighted mode, 
with simultaneous extraction of the galaxy, sky and sigma spectra.
The long GMOS slit (330 arcsec) allow us to measure the sky background in the same 
exposure as the galaxy spectrum, selected to be at the end of the slit. 

\begin{figure*}
\begin{center}	
\includegraphics[width=1.6\columnwidth]{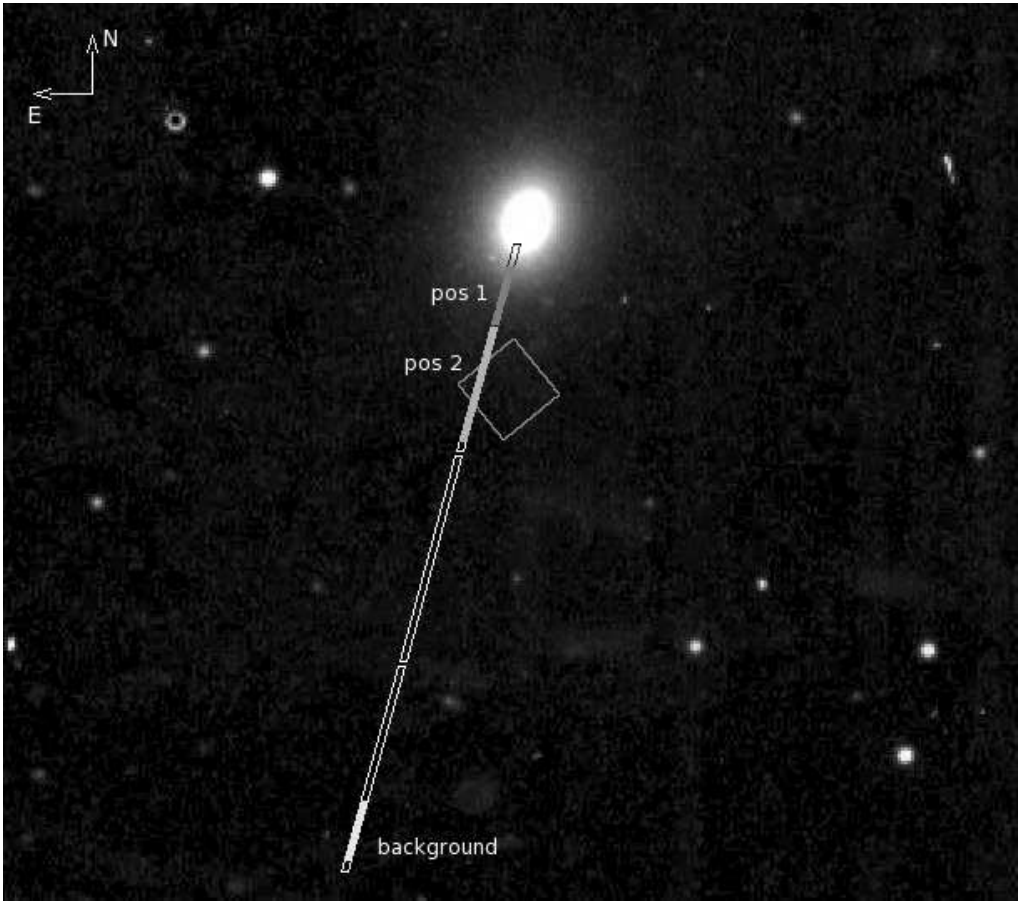} 
\hspace{1cm}
\end{center}
\caption{Digitized Sky Survey image of the field containing M32. 
The GMOS slit positioned for the long-exposures is illustrated,
with Position 1, Position 2 and sky background apertures indicated as filled regions inside the slit. 
The GMOS slit is in total 330 arcsec long. 
The square indicates the approximate position of the HST field studied by \citet{grillmair+96}.
The two discontinuities indicated in the slit correspond to the CCD gaps.}  
\protect\label{fov}
\end{figure*}

From the short exposure observations, we extracted the nuclear spectra with an aperture of 
diameter of 1.5 arcsec. From the long off-centre exposures, two apertures were extracted in each exposure: 
one covering the radius from 30 to 60 arcsec (hereafter, position 1), and another covering from 1 to 2 arcmin (hereafter, position 2). 
In Fig. \ref{fov} it is shown a Digital Sky Survey image of M32 with the GMOS slit indicated, positioned 
for the long exposures. 
The apertures extracted as positions 1 and 2 are indicated as filled regions in the
slit, as well as the region used to measure the sky background. The signal from position 1 is on average four times stronger than the background level (from eight times the sky level on the side closest to the nucleus, 
to three times in the darker side).
The signal from position 2 is on average two times stronger than the background level (from three to 1.5 times).
Position 2 was chosen
to sample the same galactic radius observed by \citet{grillmair+96} in its resolved photometric study, 
whose approximate field of view is indicated in Fig. \ref{fov} as a square.

The four spectra extracted at the central region have 
S/N\footnote{Computed as S/N$=s_{\rm galaxy}/\sqrt{s_{\rm galaxy}+s_{\rm sky}}$, 
where $s_{\rm galaxy}$ and $s_{\rm sky}$ are the galaxy and sky 
spectra, respectively.}
S/N ranging from 34 to 43 pixel$^{-1}$ at 4750\,\angstrom, 
and from 180 to 230 pixel$^{-1}$ at 7000\,\angstrom. In the case of the long off-centre exposures, 
the spectra extracted for position 1 
have S/N ranging from 50 to 80 and from 270 to 470 pixel$^{-1}$, at 4750 and 
7000\,\angstrom~respectively. The spectra for position 2
have S/N from 27 to 46 and from 152 to 267 pixel$^{-1}$ at the same wavelengths. 
The S/N of the two off-centre apertures 
drop rapidly beyond 7000\,\angstrom~due to residuals of sky subtraction.

We have not flux calibrated the spectra as no spectrophotometric standard was observed in the nights of the observations. This will not hamper our analysis, as the full 
spectrum fitting technique adopted is not sensitive to the shape of the continuum.

\section{Analysis}

\begin{figure*}
\begin{center}	
\includegraphics[width=1.8\columnwidth]{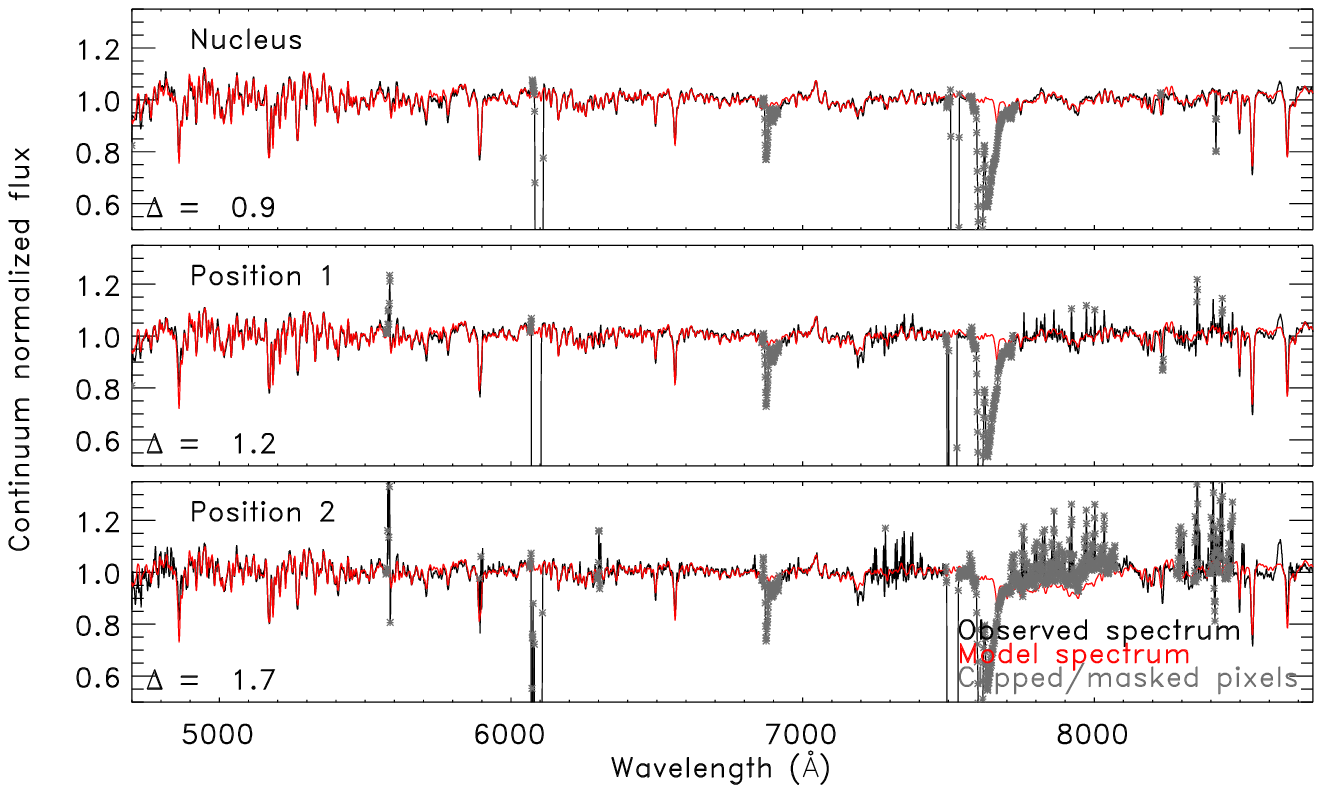} 
\hspace{1cm}
\end{center}
\caption{Observed spectra of M32 (black lines) and model spectra (red lines), for the three
different observed apertures as indicated by the labels. Masked and rejected pixels are indicated as grey 
points. The average absolute deviation $\overline{\Delta}$ of the fit (see text) is shown in each panel. Base models for these runs
are BC03.}  
\protect\label{fit_bc03}
\end{figure*}

\begin{figure*}
\begin{center}	
\includegraphics[width=1.8\columnwidth]{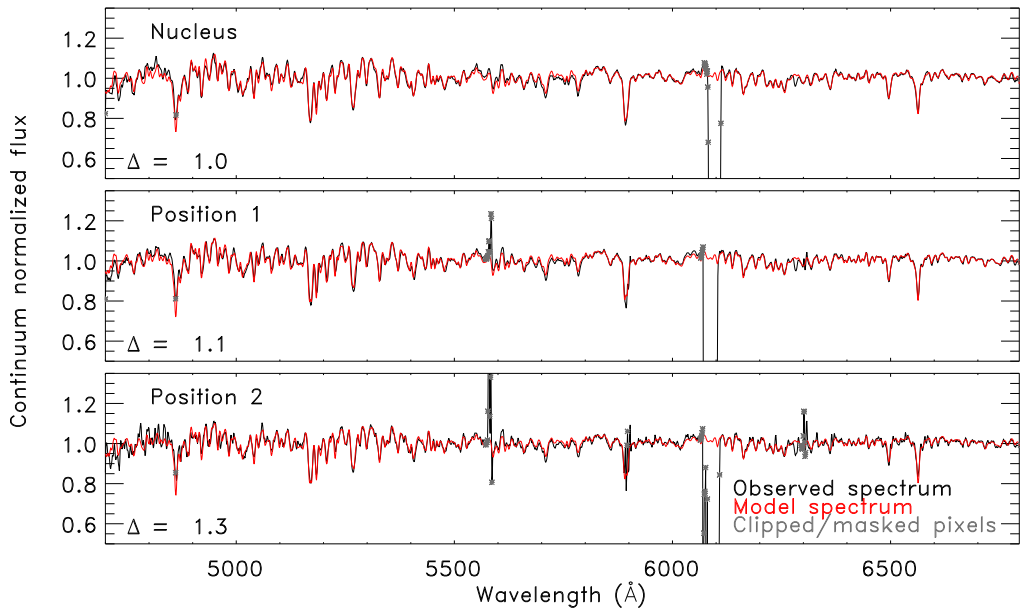} 
\hspace{1cm}
\end{center}
\caption{Same as in Figure \ref{fit_bc03}, using PEGASE-HR as base models.}  
\protect\label{fit_peg}
\end{figure*}

\begin{figure*}
\begin{center}	
\includegraphics[width=1.8\columnwidth]{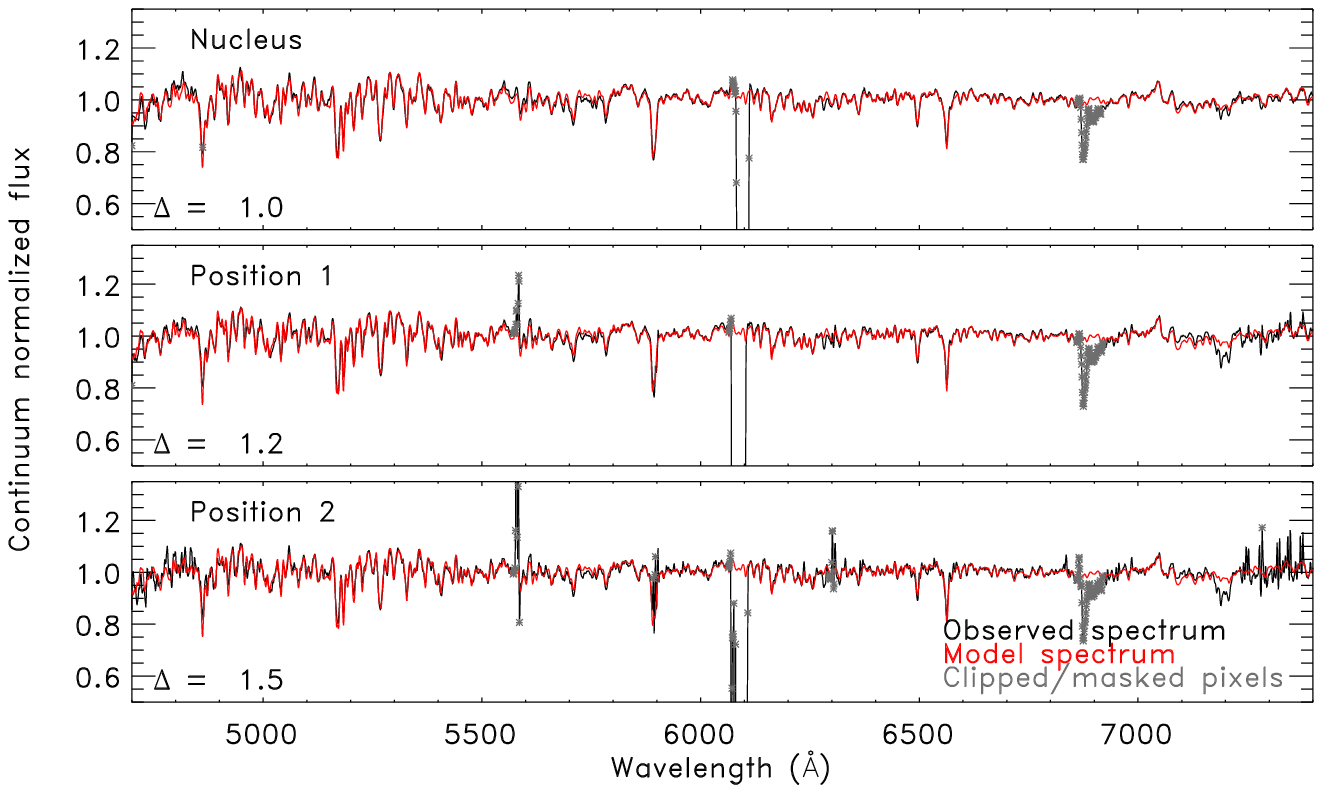} 
\hspace{1cm}
\end{center}
\caption{Same as in Figure \ref{fit_bc03}, using Vazdekis et al. as base models.}  
\protect\label{fit_vaz}
\end{figure*}

Full spectrum fitting techniques \citep[e.g.][]{panter+03,cid+05,mathis+06,ocvirk+06a,ocvirk+06b,walcher+06,koleva+08} 
have been used and validated in recent years as a powerful method for stellar population studies, 
and allows for fully exploiting the information present in our high-quality spectra of M32.

We employed a modified version of the {\sc starlight} code by
\citet{cid+05} in order to analyse the observed spectra.  {\sc
starlight} combines spectra from a user-defined base of individual
spectra in search for linear combinations that match an input observed
spectrum.  We adopt medium spectral-resolution simple SSP 
models spanning different ages and metallicities as
our spectral base (described further in the text), i.e., we describe
the data in terms of a superposition of multiple bursts of
star formation.

To explore the parameter space defined by the linear coeficients
associated with each population, the code uses a series of Markov
Chains, each starting from different locations. The likelyhood surface
is explored with a simulated annealing scheme, consisting of a
succession of steps where $\chi^2$ values are scaled by a gradually
decreasing artificial `temperature' (in analogy with Boltzmann
factors in statistical mechanics), and parameter changes are performed
with the Metropolis algorithm. In the computation of $\chi^2$ each
spectral pixel is weighted by its correspondent error, but bad pixels
(the two CCD gaps in GMOS), strong telluric absorption features and 
residuals from sky subtraction are masked out. 
Pixels which deviate by more than three times the rms
between the observed spectrum and the initial estimate for the model
spectrum are also given zero weight. The internal kinematics is
determined simultaneously with the population parameters.  The fitted
coefficients define a population vector (light fractions \textbf{x}$_j$ at a
reference wavelength of 5000\,\angstrom), which gives the combination
of SSP models in the base that best fits (smallest $\chi ^2$) the
observed spectrum. This same combination of techiques has proved
useful in several studies of the star formation history of galaxies of
different types (see, e.g., \citealt{asari+07} and the {\sc starlight}
user manual\footnote{http://www.starlight.ufsc.br/} for examples).

For the present work, the public version of the code was modified 
to allow the use of continuum-normalized spectra (as opposed to flux-calibrated spectra).
A simple moving average scheme of 200\,\angstrom~windows (excluding masked points)
was used to remove the continuum shape of both model and observed spectra.
This method fits all 
the pixels independently of the continuum, being therefore insensitive to
extinction or flux calibration errors. 
The differences between flux-calibrated versus continuum-normalized spectral fitting
have not been extensively studied yet. \citet{wolf+07} compare four methods of deriving
population parameters from integrated stellar cluster spectra, and their results favour the use
of continuum-normalized spectra, as this method performed better in 
simultaneously retrieving age and metallicity for the clusters.
We also performed a preliminary test using both versions of {\sc starlight} on the public
globular cluster spectra by \citet{schiavon+05}. We found that the two code versions yield results
that agree typically within 0.2 dex in age, and 0.3 dex in metallicity. It is possible, therefore,
that the same analysis applied on a set of flux-calibrated spectra would retrieve slightly 
different results than the ones presented here.

Spectral SSP models, adopted as base spectra, are undergoing constant revisions, 
following the improvement 
of their two main ingredients: stellar tracks and stellar spectral libraries.
As our aim is to shed some light on the composite nature of M32 stellar population, without being
biased by a particular choice of models, we opted for performing the analysis using four different sets of 
models: BC03, PEGASE-HR, Charlot \& Bruzual and 
Vazdekis et al.\footnote{http://www.iac.es/galeria/vazdekis/vazdekis\_models\_ssp\_seds.html}.
The two latter are preliminary versions of new models being developed on the basis of the work by 
BC03 and \citet{vazdekis99} respectively, using the MILES (Medium resolution INT Library of Empirical Spectra) 
empirical library \citep{sanchez-blazquez+06a,miles2}.
BC03, Charlot \& Bruzual and PEGASE-HR models were computed adopting evolutionary tracks by
\citet{bertelli+94}, and Vazdekis adopted tracks by \citet{girardi+00}. BC03 and Charlot \& Bruzual 
adopt initial mass function (IMF) from \citet{chabrier03} and PEGASE-HR and Vazdekis et al. adopt 
the IMF from \citet{salpeter55}.
For the fit procedure, we selected a range of ages between 100\,Myr and 15\,Gyr, 
spaced in log(age) = 0.25\,dex. For the
metallicites \textit{Z} we adopted the full range provided by each set of models.
The minimum wavelength of the fit is set by the observations (4700\,\angstrom), and the maximum is set by the base models 
(see Table \ref{tab_models}).

The chosen models are all based on empirical stellar spectral libraries and as such are characterized by the
abundance pattern typical of the Milky Way. Models such as those by \citet{coelho+07} can in 
principle be used with
spectral fitting techniques to extract information of the $\alpha$-enhancement in integrated light-studies.
We do not consider such models in the present study, however, 
as we understand that an analysis of the $\alpha$-enhancement 
would add an extra free parameter to an already delicate analysis. 
This will not hamper our conclusions, as \citet{worthey04} has shown
that non-solar abundance ratio effects are mild in the case of M32, and do not change the population
parameters significantly. 

In total 48 {\sc starlight} fits were performed 
(three galactic positions $\times$ four exposures for each position $\times$ four sets of base models).
Examples of the fits performed by {\sc starlight} are 
illustrated in Figures \ref{fit_bc03}, \ref{fit_peg} and \ref{fit_vaz}
for the models BC03, Pegase-HR and Vazdekis et  al. respectively. 
A measure of the quality of the fit is given by the average absolute deviation $\overline\Delta$:

\begin{equation}
\overline\Delta = \frac{1}{N} \sum_{\lambda}{| \frac{(f_{\rm model}(\lambda)-f_{\rm obs}(\lambda))}{f_{\rm obs(\lambda)}}|}
\end{equation}

\noindent where \textit{N} is the number of pixels, $f_{\rm model}$ is the fitted spectrum and $f_{\rm obs}$ is the observed spectrum. 
Typically the model spectra reproduces the observed within 2 per cent of residual flux (see Table \ref{stelpop_parameters}).
\section{Results}
\label{csp}
\begin{figure}
\begin{center}	
\includegraphics[width=\columnwidth]{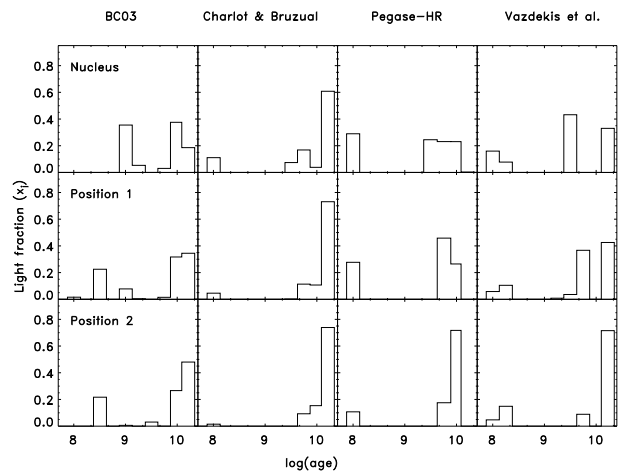} \hspace{1cm}
\end{center}
\caption{Age distribution functions obtained for each observed aperture (in rows) and each set of models (in columns), according to the labels.}
\protect\label{adf}
\end{figure}

\begin{figure}
\begin{center}	
\includegraphics[width=\columnwidth]{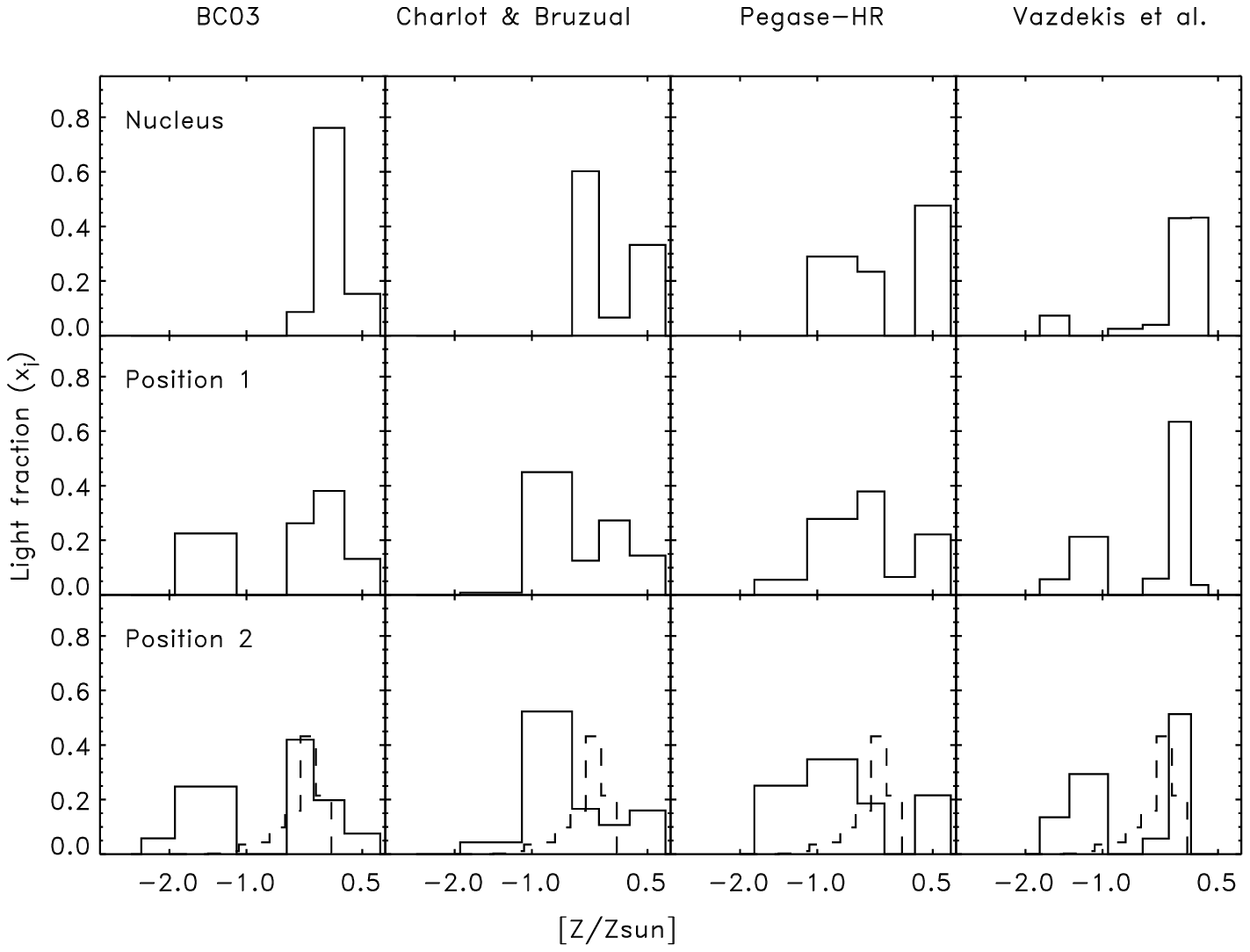} \hspace{1cm}
\end{center}
\caption{Metallicity distribution functions obtained for each observed aperture (in rows) and each set of models (in columns), according to the labels. The dashed line in the panels for Position 2 correspond to the empirical
metallicity distribution function by \citet{grillmair+96}.}  
\protect\label{mdf}
\end{figure}

The analysis was performed with each observed spectrum individually, and 
the light-fraction vectors \textbf{x}$_j$ obtained by 
{\sc starlight} were averaged over the four exposures for
each combination of base models and radius. The results of 
\textbf{x}$_j$ are qualitatively shown in Fig. \ref{popmix},
as a function of log(age) and [Z/Z$_{\odot}$], and in tabular form in the Appendix A.
The projections of the averaged \textbf{x}$_j$ in the log(age) and $Z$ planes give the age distribution
functions (ADFs) and metallicity distribution functions (MDFs), 
shown in Figs \ref{adf} and \ref{mdf}, respectively.

\begin{figure*}
\begin{center}	
\includegraphics[width=14cm]{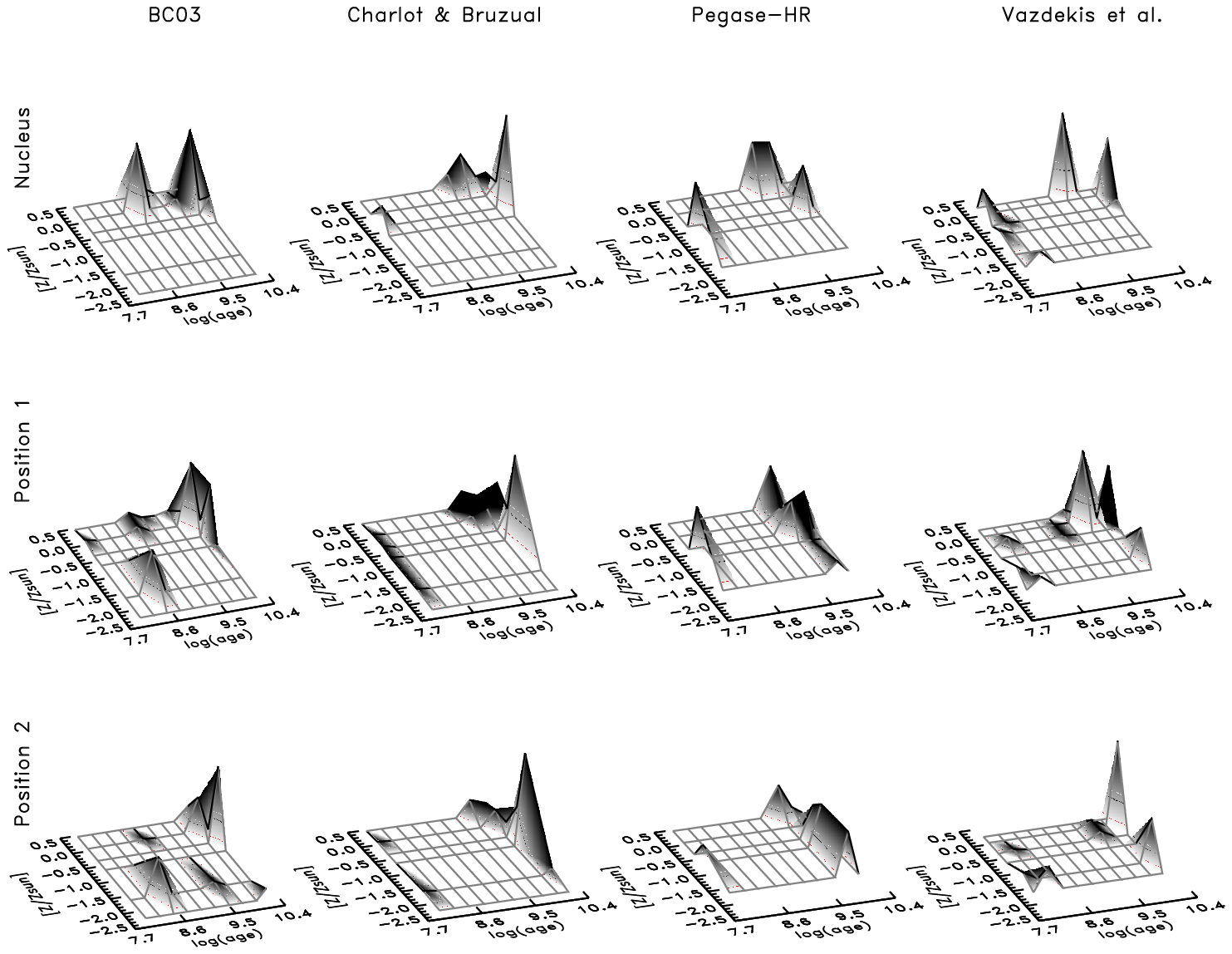} \hspace{1cm}
\end{center}
\caption{Representation of the light-fraction vectors x$_j$ as a function of log(age) and [Z/Z$_{\odot}$].
The results for the nucleus, positions 1 and 2 are shown at the top, middle and bottom rows, respectively.
Each column shows the results for a different set of base models, as indicated at the top of the figure. The
light-fractions are given in tabular form in the appendix.}  
\protect\label{popmix}
\end{figure*}

It is worth remembering that the inversion from an observed spectrum to a set of 
parameters is a complex and degenerate problem, and we should avoid 
overdetailed descriptions of the population mixture \citep[][]{ocvirk+06a}.
With that caveat in mind, we draw the 
following conclusions.

\subsection{The distribution of ages} 

The age distribution has two to three peaks, depending on the base model.
Signal of an ancient population [log(age) $\ge$ 10] 
is retrieved in all positions.
Previous evidence of the ancient population were obtained from photometric studies, 
both for the central region \citep{brown+00,brown+08} and the outer reaches of 
the galaxy, $\sim$ 3\,arcmin away from the centre \citep[][]{alonsogarcia+04}. 

Intermediate-age populations [log(age) between 9.0 and 9.9] are also always retrieved.
Most of the previous spectral studies find a mean intermediate age for the population of M32 
(e.g. \citealt{rose+05} find a SSP age between 4 and 7\,Gyr, depending on the radius), and
the existence of such population also has been supported by the photometric studies  
by \citet{davidge_jensen07} and references therein.

In many cases, the inversion algorithm retrieves small amounts of a young population [log(age) $<$ 9.0]),  
typically stronger in the runs with PEGASE-HR models.
\citet{rose+05} noted that a population
in M32 cannot be any younger than 2\,Gyr, or it would produce a strong signature in the Ca II 
index. As our spectrum does not extend blueward than 
4800\,\angstrom, the Ca II region is not considered in the fit.
The determination of ages in integrated spectra 
may be a difficult problem, as extended horizontal 
branch morphologies and/or blue stragglers, unaccounted for 
in the SSP models, can mimic young ages \citep{pacheco_barbuy95,maraston_thomas00,trager+05}. 
Interpretation of integrated globular cluster spectra from \citet{schiavon+05} with
STECMAP \citep[STEllar Content via Maximum A Posteriori;][]{ocvirk+06a} supports the idea that blue HB stars in old metal-poor populations
might appear as a spurious young burst of star formation (\citealt{koleva+07bproc,koleva+08}, Ocvirk, private communication).
Hot HB stars in the central regions
of M32 have been found by \citet{brown+00}, and 
thus it is possible that a least a fraction of the young population 
retrieved here is a signal of these hot old stars.

\subsection{The distribution of abundances}

The MDFs functions are smoother than the age distributions, with one or two peaks.
In most of the cases, the peak of the metallicity distribution is centred at sub-solar values.

Position 2, the farthest from the nucleus, samples the same radius that was observed by
\citet{grillmair+96}. The metallicity distribution obtained in that work is overplotted to our
distributions as a dashed line in Fig. \ref{mdf}. 
Our results are in agreement with the conclusion by Grillmair et al. that the metallicity
distribution in M32 is broad. On the other hand, they obtain a distribution which is poor in 
stars with [Z/Z$_{\odot}$] $\leq$ $-$1. This is, in contrast, 
an important fraction in our mixtures regardless of
the base models adopted.

Several effects may be hampering the comparison between the photometric and the spectroscopic
metallicity distributions.
The MDF by \citet{grillmair+96} is based on the colour distribution of red giant branch (RGB) stars, 
and as such it is affected by observational errors and age effects (their
distribution assumes a constant age of 10\,Gyr). 
Furthermore, the set of underlying isochrones of the spectral models used by the present work 
\citep{bertelli+94,girardi+00} is different from the isochrones used for the star counts
in Grillmair et al. \citep{worthey+94}.
Moreover, the MDF by Grillmair et al. is in the form of 
number of stars per \textit{Z} bin, i.e., \textit{N}(Z)dZ, while ours is a light-fraction MDF, 
L$_{\rm tot}$(Z)dZ. 
In a general case where L$_{\rm tot}$ is not constant with \textit{Z}, then \textit{N}(Z) and \textit{L}(Z) are not the same, 
and one should not expect a perfect match. 
As pointed out by the anonymous referee, one might not expect 
an MDF derived from spectroscopic data to
match an MDF derived from photometric data unless biases and observational errors are properly 
taken into account.

In the mixtures retrieved by our analysis, the metal-poor component corresponds to a population
younger than the metal-rich one (according to Fig. \ref{popmix} and Tables A1$-$A9). 
It would be particularly puzzling if this trend is real, as it is in general
expected that younger populations are formed from interstelar material
already processed by older populations, and thus should be more metal-rich. 
On the other hand, in some cases this younger metal-poor population appears at the edges
of the parameter space covered by the models, what might be an indication of an artefact effect.
 
\subsection{Mean population parameters}
\label{ssp}

The majority of the previous studies of M32 stellar populations publish 
the age and metallicities derived by the comparison to SSP models.
Contrary to what is sometimes assumed, the SSP-equivalent ages and metallicities 
are not the same as the 
luminosity-weighted age and metallicites
\citep[see e.g.][]{trager+00b,walcher+06,trager_somerville09}.

We computed the luminosity-weighted parameters of our 
population mixtures by adopting:

\begin{center}
\begin{equation}
\langle\log({\rm age})\rangle = \sum_{j}{x_j \cdot \log({\rm age})_j}
\end{equation}
\end{center}

\noindent and

\begin{center}
\begin{equation}
\langle[Z/Z_{\odot}]\rangle = \sum_{j}{x_j \cdot [Z/Z_{\odot}]_j}
\end{equation}
\end{center}

\noindent where the vector \textbf{x}$_j$ gives the normalized light-fraction of the $j$th SSP
component of the model fit ($\sum_{j}{\textbf{x}_j}=1$). 
The SSP-equivalent parameters, on the other hand, are the age and metallicity 
of the $j$th SSP which best fits the observed spectrum (smallest $\overline\Delta$). 

We list both the luminosity-weighted and SSP-equivalent parameters in Table \ref{stelpop_parameters},
derived independently from each of the four exposures.
It can be seen that the difference from adopting distinct base models is larger than
the variance of the {\sc starlight} runs on the four exposures of each position.
 
\begin{table*}
\caption{Light-weighted mean parameters and SSP-equivalent parameters obtained from the fits. 
Results are given for individual observed spectra.}
\label{stelpop_parameters}
\begin{center}
\begin{tabular}{lc|ccc|ccc}
\hline
 &      &     \multicolumn{3}{|c|}{Light-weighted parameters} & \multicolumn{3}{|c|}{Best SSP fit parameters}\\
\hline
Region    & Obs \# & Age[Gyr]   & [Z/Z$_\odot$] & $\overline\Delta$[\%] &  Age[Gyr]   & [Z/Z$_\odot$]  & $\overline\Delta$ [\%] \\
\noalign{\smallskip} \hline \multicolumn{8}{c}{{\it BC03} }  \\
Nucleus        & 1       &      4.0 &   0.12&  0.93&      3.2 &  0.07&  0.94\\
               & 2       &      4.5 &   0.09&  0.98&      3.2 &  0.07&  0.98\\
               & 3       &      4.6 &   0.07&  0.97&      3.2 &  0.07&  0.97\\
               & 4       &      4.1 &   0.12&  1.00&      3.2 &  0.07 & 1.00 \\
Position 1     & 1       &      5.4 &  -0.13 & 1.14 &    10.0 & -0.33 & 1.18 \\
               & 2       &      3.7 &  -0.38 & 1.09 &    15.0 & -0.33&  1.14\\
               & 3       &      3.8 &  -0.45 & 1.14 &    15.0 & -0.33&  1.20\\
               & 4       &      3.6 &  -0.49 & 1.19 &    15.0 & -0.33 & 1.24 \\
Position 2     & 1       &      7.8 &  -0.55 & 2.11 &    10.0 & -0.33 & 2.13 \\
               & 2       &      4.6 &  -0.65 & 1.51 &     5.5 & -0.33 & 1.52 \\
               & 3       &      5.6 &  -0.66 & 1.54 &    10.0 & -0.33 & 1.57 \\
               & 4       &      4.6 &  -0.63 & 1.65 &    10.0 & -0.33 & 1.65 \\
\noalign{\smallskip} \hline  \multicolumn{8}{c}{{\it Charlot \& Bruzual } }  \\
Nucleus        & 1       &       6.1&  -0.05& 0.89&      5.5&  0.07 &   0.95 \\
               & 2       &       6.4&  -0.03& 0.94&      5.5&  0.07 &   0.99 \\
               & 3       &       6.8&  -0.02& 0.93&      5.5&  0.07 &   0.96 \\
               & 4       &       6.3&  -0.04& 0.96&      5.5&  0.07 &   1.00 \\
Position 1     & 1       &       6.8&  -0.21& 1.02&     10.0 &-0.33 &   1.04 \\
               & 2       &      10.0 & -0.25& 1.05 &    10.0 &-0.33 &   1.09 \\
               & 3       &      11.7 & -0.26& 1.01 &    10.0 &-0.33 &   1.04 \\
               & 4       &      13.3 & -0.28& 1.03 &    10.0& -0.33 &   1.05 \\
Position 2     & 1       &      11.5&  -0.49& 1.93&     10.0& -0.33 &   1.95 \\
               & 2       &      10.8 & -0.30& 1.33 &     5.5 &-0.33 &   1.35 \\
               & 3       &      12.6 & -0.33& 1.37 &    10.0& -0.33 &   1.40 \\
               & 4       &      12.8&  -0.37& 1.45&      5.5& -0.33&   1.49\\
\noalign{\smallskip} \hline \multicolumn{8}{c}{{\it Pegase-HR}  } \\
Nucleus        & 1       &      1.7&  -0.03& 1.04&    3.0 &  0.07&  1.04\\
               & 2       &      1.5&  -0.04& 1.10&    3.0 &  0.07&  1.11\\
               & 3       &      2.0&  -0.03& 1.07&    3.0 &  0.07&  1.07\\
               & 4       &      1.9&  -0.03& 1.10&    3.0 &  0.07 & 1.10 \\
Position 1     & 1       &      3.6&  -0.37& 1.13&    6.0 & -0.33 & 1.18 \\
               & 2       &      1.7&  -0.25& 1.11&    6.0 & -0.33 & 1.13 \\
               & 3       &      1.9&  -0.25& 1.10&    6.0 & -0.33 & 1.12 \\
               & 4       &      2.0&  -0.25& 1.13&    6.0 & -0.33 & 1.15 \\
Position 2     & 1       &      8.8&  -0.78& 2.06&    3.0 & -0.33 & 2.10 \\
               & 2       &      4.2&  -0.49& 1.23&    3.0 & -0.33 & 1.27 \\
               & 3       &      4.7&  -0.52& 1.28&    3.0 & -0.33 & 1.32 \\
               & 4       &      5.6&  -0.55& 1.31&    3.0 & -0.33 & 1.35 \\
\noalign{\smallskip} \hline \multicolumn{8}{c}{{\it Vazdekis et al.}  }  \\
Nucleus        & 1       &      2.3&  -0.02&  0.97&     3.2 &  0.05 & 0.97 \\
               & 2       &      2.4&   0.02&  1.01&     3.2 &  0.05 & 1.02 \\
               & 3       &      2.7&  -0.08&  1.01&     3.2 &  0.05 & 1.01 \\
               & 4       &      2.5&  -0.01&  1.04&     3.2 &  0.05 & 1.04 \\
Position 1     & 1       &      3.9&  -0.37&  1.07&     5.6 & -0.33 & 1.11 \\
               & 2       &      4.5&  -0.30&  1.13&     3.2 &  0.05 & 1.17 \\
               & 3       &      5.0&  -0.30&  1.15&     3.2 &  0.05 & 1.18 \\
               & 4       &      5.3&  -0.38&  1.23&     5.6 & -0.33 & 1.26 \\
Position 2     & 1       &      4.3 & -0.68 & 1.95 &    5.6 & -0.33 & 1.99 \\
               & 2       &      5.2 & -0.53 & 1.39 &    3.2 & -0.33 & 1.38 \\
               & 3       &      6.5 & -0.52 & 1.41 &    5.6 & -0.33 & 1.44 \\
               & 4       &      8.0 & -0.58 & 1.52 &   10.0 & -0.63 & 1.52 \\
\hline
\end{tabular}
\end{center}
\end{table*}

\subsection{Evidence for gradients ?}

Studies of radial colour and line strength gradients have led to discordant results in recent years 
\citep[see discussion in][ and references therein]{rose+05}.
In the present work, we opted for maximising the S/N 
\citep[a critical issue in spectral fitting technique, see e.g.][]{koleva+07aproc}, and thus 
we have not split our spectra in many apertures to perform a detailed gradient study.
In general terms, it is possible to identify in our analysis a shift in the distribution
of populations, in the sense that the fraction coming from an old population is higher in the outer 
regions than in the nucleus.
In order to statistically address this issue, 
in Table \ref{average_parameters} we show the light-weigthed parameters from 
Table \ref{stelpop_parameters} averaged on the four exposures, with their corresponding
1\,$\sigma$ uncertainties. A trend of higher ages and lower
metallicities as one moves from the nucleus to position 2 can be seen in all cases,
even though the gradient slopes are model-dependent.

\begin{table}
\caption{Average luminosity-weighted parameters.}
\label{average_parameters}
\begin{center}
\begin{tabular}{lccc}
\hline
Region &  Age (Gyr) & [Z/Z$_\odot$]\\
\noalign{\smallskip} \hline \multicolumn{3}{c}{{\it BC03} }  \\
Nucleus        & 4.3 $\pm$ 0.3 &      0.1 $\pm$ 0.1 \\
Position 1     & 4.1 $\pm$ 0.9 &     -0.4 $\pm$ 0.2 \\
Position 2     & 5.7 $\pm$ 1.5 &     -0.6 $\pm$ 0.1 \\
\noalign{\smallskip} \hline \multicolumn{3}{c}{{\it Charlot \& Bruzual} }  \\
Nucleus        & 6.4 $\pm$ 0.3 &      0.0 $\pm$ 0.1 \\
Position 1     &10.4 $\pm$ 2.8 &     -0.2 $\pm$ 0.1 \\
Position 2     &11.9 $\pm$ 0.9 &     -0.4 $\pm$ 0.1 \\
\noalign{\smallskip} \hline \multicolumn{3}{c}{{\it Pegase-HR} }  \\
Nucleus        & 1.8 $\pm$ 0.2 &      0.0 $\pm$ 0.1 \\
Position 1     & 2.3 $\pm$ 0.9 &     -0.3 $\pm$ 0.1 \\
Position 2     & 5.8 $\pm$ 2.1 &     -0.6 $\pm$ 0.1 \\
\noalign{\smallskip} \hline \multicolumn{3}{c}{{\it Vazdekis et al.} }  \\
Nucleus        & 2.5 $\pm$ 0.2 &      0.0 $\pm$ 0.1 \\
Position 1     & 4.7 $\pm$ 0.6 &     -0.3 $\pm$ 0.1 \\
Position 2     & 6.0 $\pm$ 1.6 &     -0.6 $\pm$ 0.1 \\
\hline
\end{tabular}
\end{center}
\end{table}

These findings supports one of the scenarios discussed by \citet{rose+05},
in which there is a shifting balance between the populations with radii. 
A recent burst of star-formation on top of an ancient population
would appear as a higher concentration of a younger 
and more metal-rich population 
in the nucleus. This scenario would be a natural
result of the model proposed by \citet{bekki+01},
in which centralised star formation in M32 is triggered by the tidal field 
of M31. On the other hand, \textit{HST} near-ultraviolet (UV) imaging of M32 by 
\citet{brown+98} reveals that the resolved UV-bright stars are 
more centrally concentrated than the underlying unresolved UV 
sources. Thus, it is also possible that the trend we see is a
changing influence of a relatively small hot population prevalent
in the nucleus. As the models do not yet consider hot stars in
old populations, it is not possible with the present
analysis to distinguish if it is the hot-old stars or the hot-young stars 
(or a combination of both) which is driving this balance 
shift.\footnote{\textit{Note added in proof}: 
An alternative scenario to explain our findings 
is the one proposed recently by Kormendy et al. (2009). 
In this formation scenario, the young light in the
centre would be the result of a starburst in the latest 
dissipative merger that made the galaxy. 
There is a gradual change as radius increases, towards
larger contributions from older stars that formed before the most
recent merger(s). In this scenario, the gradient would be therefore intrinsic to the 
formation process of M32, and not triggered by M31.}

\subsection{Dependence of the results with wavelength range}

For the main fitting runs we opted for maximizing the use of information and 
thus to perform the fitting on the total wavelength range allowed by each set
of spectral models. We additionally performed tests fitting only the region 
which is common to all base models (4700$-$6800\,\angstrom) and a limited region
encompassing the most commonly used Lick/IDS
(Lick Observatory Image Dissector Scanner)
indices (4828$-$5363\,\angstrom, from H$\beta$ to Fe5335).

The results fitting only the region 4700$-$6800\,\angstrom~are not strongly different 
from those obtained when fitting the whole range, except that lower weight is given to the 
old populations, and thus mean ages which are roughly 1\,Gyr younger are derived.
In the case of fitting the 4828$-$5363\,\angstrom~region, 
the ages retrieved are slightly older in the nucleus but younger in the outer regions, and thus
the gradient is flatter. The MDFs do not change significantly, 
except that the mean metallicity is slightly higher when 
only the 4828$-$5363\,\angstrom~range is fitted. 

The dependence of the results on wavelength range is an expected behavior
for galaxies with star formation histories that are not well represented by a 
single burst \citep[e.g.][]{sanchez-blazquez+06b,schiavon07}.
Therefore, we favor the use of wide wavelength ranges in analyses that aim at
studying composite stellar populations.

\section{Summary and Conclusions}

We obtained high S/N integrated spectra of the galaxy M32 with the 
GMOS at Gemini-North, out to 2 arcmin (3.6\,$r_{\rm eff}$) from the centre of the galaxy. 
We performed an analysis of the 
composite stellar population in M32 using {\sc starlight} by \citet{cid+05}, 
a technique of full spectrum
fitting, and adopting different sets of medium spectral resolution 
stellar population models: \citet{BC03}, \citet{PEGASE-HR}, Charlot \& Bruzual (in prep.) and 
Vazdekis et al. (in prep.). Three different regions of the galaxy were analysed
separately, the nuclear region, a region from 30 to 60\,arcsec, and a region from 1 to 2\,arcmin.

We can find mixtures of populations that reproduce our observed spectra within 
2 per cent of residual flux. We find that in the nuclear region, 23$-$65\,per cent
(depending on the model used) of the light 
at 5000\,\angstrom~comes from an ancient population [older than log(age) = 10], 
and intermediate-age populations [log(age) between 9 and 9.9] contribute to 
25$-$48\,per cent of the light. 
Beyond 1\,$r_{\rm eff}$ ($r_{\rm eff} \sim 33"$), the light coming from the
ancient population increases (27$-$90\,per cent) and the contribution from intermediate-age
populations decreases (4$-$41\,per cent). 
There is a signal of a young-population [log(age) younger than 9.0] that contributes
up to 30\,per cent. 
As no evidence of such young population has been found by 
previous photometric and spectroscopic studies, it is likely that at
least part of this populations is a spectral signal of 
the hot HB stars identified by \citet{brown+00}. These findings suggest that the lack of 
spectral population models accouting for these hot old stars may set a 
constraint in our ability to identify the detailed age distribution of a galaxy
from its integrated spectrum.

The metallicity distribution is peaked at subsolar values, 
and we retrieve a non-negligible
contribution from a metal-poor population with [Z/Z$_{\odot}$]~$\leq$~$-$1.0. 
Such metal-poor population has not been identified in the CMD analysis such as \citet{grillmair+96}.
As our position 2 samples the same radius as the analysis by Grillmair et al., 
this disagreement between integrated light studies and resolved
stellar studies has yet to be understood.

As the main result of the present analysis, we propose that an ancient and an 
intermediate-age population co-exist in M32,
in agreement with the photometric evidence found by 
e.g. \citet{brown+00,alonsogarcia+04} (for an old population) and 
\citet{davidge_jensen07} (for an intermediate age population). 
Moreover, we propose that the balance between these two populations 
change from the nucleus to the halo (outside 1\,$r_{\rm eff}$) 
in the sense that the contribution from the intermediate population 
is larger at the nuclear region.

More detailed descriptions of the age and metallicity distributions in M32
may have to wait for future stellar population models, 
as the balance of populations that fit our high-quality integrated spectra
is model-dependent.

\vspace{1cm}

{\it Acknowledgments: }
PC acknowledges the support of the European Community under
a Marie Curie International Incoming Fellowship (6th Framework
Programme, FP6), and of FAPESP (Funda\c c\~ao de Amparo \`a Pesquisa
do Estado de S\~ao Paulo) under the Post-Doc Fellowship No 05/03840-3.
This research has also been supported by FAPESP Projeto Tem{\'a}tico No 06/56213-9.
PC would like to thank G. Bruzual, S. Charlot 
and A. Vazdekis for kindly providing access to their new models
prior to publication. 
The authors also would like to thank the anonymous referee for his/her
much valuable comments.

\label{lastpage}
\bibliographystyle{mn2e}

\appendix

\section{Results}
Population mixtures obtained for each set of SSP models, averaged on the four exposures
(Tables A1$-$A12).

\clearpage

\begin{table}
\caption{Light percentage fraction as a function of age and Z for the nuclear region, obtained with BC03 models. The values for [Z/Z$_{\sun}$] are given in the column headers, 
log(age) is given in the first column. Adopted: Z$_{\sun}$ = 0.017.}
\label{tab_bc03_center}
\begin{center}
\begin{tabular}{lcccccc}
\hline
\multicolumn{1}{r}{[Z/H]}  &   $-$2.23 &   $-$1.63 &   $-$0.63 &   $-$0.33 &    0.07 &    0.47 \\
log(age) \\
\hline
 8.00 &      0 &      0 &      0 &      0 &      0 &      0 \\
 8.25 &      0 &      0 &      0 &      0 &      0 &      0 \\
 8.50 &      0 &      0 &      0 &      0 &      0 &      0 \\
 8.75 &      0 &      0 &      0 &      0 &      0 &      0 \\
 9.00 &      0 &      0 &      0 &      0 &     36 &      0 \\
 9.25 &      0 &      0 &      0 &      0 &      0 &      5 \\
 9.50 &      0 &      0 &      0 &      0 &      0 &      0 \\
 9.75 &      0 &      0 &      0 &      0 &      3 &      0 \\
10.00 &      0 &      0 &      0 &      0 &     38 &      0 \\
10.18 &      0 &      0 &      0 &      9 &      0 &     10 \\
\hline
\end{tabular}
\end{center}
\end{table}

\begin{table}
\caption{Same as in \ref{tab_bc03_center}, but for position 1 and BC03 models.}
\begin{center}
\begin{tabular}{lcccccc}
\hline
\multicolumn{1}{r}{[Z/H]}  &   $-$2.23 &   $-$1.63 &   $-$0.63 &   $-$0.33 &    0.07 &    0.47 \\
log(age) \\
\hline
 8.00 &    0 &      0 &      0 &      0 &      2 &      0  \\
 8.25 &    0 &      0 &      0 &      0 &      0 &      0  \\
 8.50 &    0 &     23 &      0 &      0 &      0 &      0  \\
 8.75 &    0 &      0 &      0 &      0 &      0 &      0  \\
 9.00 &    0 &      0 &      0 &      0 &      4 &      4  \\
 9.25 &    0 &      0 &      0 &      0 &      0 &      0  \\
 9.50 &    0 &      0 &      0 &      0 &      0 &      0  \\
 9.75 &    0 &      0 &      0 &      0 &      0 &      1  \\
10.00 &    0 &      0 &      0 &      0 &     32 &      0  \\
10.18 &    0 &      0 &      0 &     26 &      1 &      7  \\
\hline
\end{tabular}
\end{center}
\end{table}

\begin{table}
\caption{Same as in \ref{tab_bc03_center}, but for position 2 and BC03 models.}
\begin{center}
\begin{tabular}{lcccccc}
\hline
\multicolumn{1}{r}{[Z/H]}  &   $-$2.23 &   $-$1.63 &   $-$0.63 &   $-$0.33 &    0.07 &    0.47 \\
log(age) \\
\hline
 8.00 &    0 &      0 &      0 &      0 &      0 &      0  \\
 8.25 &    0 &      0 &      0 &      0 &      0 &      0  \\
 8.50 &    0 &     21 &      0 &      0 &      0 &      0  \\
 8.75 &    0 &      0 &      0 &      0 &      0 &      0  \\
 9.00 &    0 &      0 &      0 &      0 &      1 &      0  \\
 9.25 &    0 &      0 &      0 &      0 &      0 &      0  \\
 9.50 &    0 &      3 &      0 &      0 &      0 &      0  \\
 9.75 &    0 &      0 &      0 &      0 &      0 &      0  \\
10.00 &    0 &      0 &      0 &      0 &     19 &      8  \\
10.18 &    6 &      0 &      0 &     42 &      0 &      0  \\
\hline
\end{tabular}
\end{center}
\end{table}

\begin{table}
\caption{Same as in \ref{tab_bc03_center}, but for the nuclear region and Charlot \& Bruzual models.}
\begin{center}
\begin{tabular}{lcccccc}
\hline
\multicolumn{1}{r}{[Z/H]}  &   $-$2.23 &   $-$1.63 &   $-$0.63 &   $-$0.33 &    0.07 &    0.47 \\
log(age) \\
\hline
 8.00 &      0 &      0 &      0 &     11 &      0 &      0\\
 8.25 &      0 &      0 &      0 &      0 &      0 &      0\\
 8.50 &      0 &      0 &      0 &      0 &      0 &      0\\
 8.75 &      0 &      0 &      0 &      0 &      0 &      0\\
 9.00 &      0 &      0 &      0 &      0 &      0 &      0\\
 9.25 &      0 &      0 &      0 &      0 &      0 &      0\\
 9.50 &      0 &      0 &      0 &      0 &      0 &      8\\
 9.75 &      0 &      0 &      0 &      0 &      0 &     17\\
10.00 &      0 &      0 &      0 &      0 &      0 &      4\\
10.18 &      0 &      0 &      0 &     49 &      7 &      5\\
\hline
\end{tabular}
\end{center}
\end{table}

\begin{table}
\caption{Same as in \ref{tab_bc03_center}, but for position 1 and Charlot \& Bruzual models.}
\begin{center}
\begin{tabular}{lcccccc}
\hline
\multicolumn{1}{r}{[Z/H]}  &   $-$2.23 &   $-$1.63 &   $-$0.63 &   $-$0.33 &    0.07 &    0.47 \\
log(age) \\
\hline
 8.00 &    0 &      1 &      1 &      2 &      1 &      0\\
 8.25 &    0 &      0 &      0 &      0 &      0 &      0\\
 8.50 &    0 &      0 &      0 &      0 &      0 &      0\\
 8.75 &    0 &      0 &      0 &      0 &      0 &      0\\
 9.00 &    0 &      0 &      0 &      0 &      0 &      0\\
 9.25 &    0 &      0 &      0 &      0 &      0 &      0\\
 9.50 &    0 &      0 &      0 &      0 &      0 &      0\\
 9.75 &    0 &      0 &      0 &      0 &      2 &      9\\
10.00 &    0 &      0 &      0 &      0 &      5 &      5\\
10.18 &    0 &      0 &     44 &     11 &     18 &      0\\
\hline
\end{tabular}
\end{center}
\end{table}

\begin{table}
\caption{Same as in \ref{tab_bc03_center}, but for position 2 and Charlot \& Bruzual models.}
\begin{center}
\begin{tabular}{lcccccc}
\hline
\multicolumn{1}{r}{[Z/H]}  &   $-$2.23 &   $-$1.63 &   $-$0.63 &   $-$0.33 &    0.07 &    0.47 \\
log(age) \\
\hline
 8.00 &    0 &      1 &      0 &      0 &      1 &      0  \\
 8.25 &    0 &      0 &      0 &      0 &      0 &      0  \\
 8.50 &    0 &      0 &      0 &      0 &      0 &      0  \\
 8.75 &    0 &      0 &      0 &      0 &      0 &      0  \\
 9.00 &    0 &      0 &      0 &      0 &      0 &      0  \\
 9.25 &    0 &      0 &      0 &      0 &      0 &      0  \\
 9.50 &    0 &      0 &      0 &      0 &      0 &      0  \\
 9.75 &    0 &      0 &      0 &      0 &      0 &      9  \\
10.00 &    0 &      0 &      0 &      0 &      9 &      7  \\
10.18 &    0 &      4 &     52 &     17 &      1 &      0  \\
\hline
\end{tabular}
\end{center}
\end{table}

\begin{table}
\caption{Same as in \ref{tab_bc03_center}, but for the nuclear region and PEGASE-HR models.}
\begin{center}
\begin{tabular}{lccccc}
\hline
\multicolumn{1}{r}{[Z/H]}  &   $-$1.63 &   $-$0.63 &   $-$0.33 &    0.07 &    0.47 \\
log(age) \\
\hline
 8.00 &     0 &     29 &      0 &      0 &      0\\
 8.25 &     0 &      0 &      0 &      0 &      0\\
 8.50 &     0 &      0 &      0 &      0 &      0\\
 8.75 &     0 &      0 &      0 &      0 &      0\\
 9.00 &     0 &      0 &      0 &      0 &      0\\
 9.25 &     0 &      0 &      0 &      0 &      0\\
 9.50 &     0 &      0 &      0 &      0 &     25\\
 9.75 &     0 &      0 &      0 &      0 &     23\\
10.00 &     0 &      0 &     23 &      0 &      0\\
10.18 &     0 &      0 &      0 &      0 &      0\\
\hline
\end{tabular}
\end{center}
\end{table}

\begin{table}
\caption{Same as in \ref{tab_bc03_center}, but for position 1 and PEGASE-HR models.}
\begin{center}
\begin{tabular}{lccccc}
\hline
\multicolumn{1}{r}{[Z/H]}  &   $-$1.63 &   $-$0.63 &   $-$0.33 &    0.07 &    0.47 \\
log(age) \\
\hline
 8.00 &     0 &     28 &      3 &      0 &      0\\
 8.25 &     0 &      0 &      0 &      0 &      0\\
 8.50 &     0 &      0 &      0 &      0 &      0\\
 8.75 &     0 &      0 &      0 &      0 &      0\\
 9.00 &     0 &      0 &      0 &      0 &      0\\
 9.25 &     0 &      0 &      0 &      0 &      0\\
 9.50 &     0 &      0 &      0 &      0 &      0\\
 9.75 &     0 &      0 &     17 &      6 &     22\\
10.00 &     6 &      0 &     21 &      0 &      0\\
10.18 &     0 &      0 &      0 &      0 &      0\\
\hline
\end{tabular}
\end{center}
\end{table}

\begin{table}
\caption{Same as in \ref{tab_bc03_center}, but for position 2 and PEGASE-HR models.}
\begin{center}
\begin{tabular}{lcccccc}
\hline
\multicolumn{1}{r}{[Z/H]}  &   $-$1.63 &   $-$0.63 &   $-$0.33 &    0.07 &    0.47 \\
log(age) \\
\hline
 8.00 &     0 &     11 &      0 &      0 &      0  \\
 8.25 &     0 &      0 &      0 &      0 &      0  \\
 8.50 &     0 &      0 &      0 &      0 &      0  \\
 8.75 &     0 &      0 &      0 &      0 &      0  \\
 9.00 &     0 &      0 &      0 &      0 &      0  \\
 9.25 &     0 &      0 &      0 &      0 &      0  \\
 9.50 &     0 &      0 &      0 &      0 &      0  \\
 9.75 &     0 &      0 &      0 &      0 &     17  \\
10.00 &    25 &     24 &     19 &      0 &      4  \\
10.18 &     0 &      0 &      0 &      0 &      0  \\
\hline
\end{tabular}
\end{center}
\end{table}
 
\begin{table}
\caption{Same as in \ref{tab_bc03_center}, but for the nuclear region and Vazdekis et al. models.}
\begin{center}
\begin{tabular}{lcccccc}
\hline
\multicolumn{1}{r}{[Z/H]}  &   $-$1.63 &   $-$1.23 &   $-$0.63 &   $-$0.33 &    0.05 &    0.25 \\
log(age) \\
\hline
 8.00 &    0 &      0 &      3 &      0 &     13 &      0  \\
 8.25 &    7 &      0 &      0 &      0 &      0 &      0  \\
 8.50 &    0 &      0 &      0 &      0 &      0 &      0  \\
 8.75 &    0 &      0 &      0 &      0 &      0 &      0  \\
 9.00 &    0 &      0 &      0 &      0 &      0 &      0  \\
 9.25 &    0 &      0 &      0 &      0 &      0 &      0  \\
 9.50 &    0 &      0 &      0 &      0 &      0 &     43  \\
 9.75 &    0 &      0 &      0 &      0 &      0 &      0  \\
10.00 &    0 &      0 &      0 &      0 &      0 &      0  \\
10.18 &    0 &      0 &      0 &      4 &     29 &      0  \\
\hline
\end{tabular}
\end{center}
\end{table}

\begin{table}
\caption{Same as in \ref{tab_bc03_center}, but for position 1 and Vazdekis et al. models.}
\begin{center}
\begin{tabular}{lcccccc}
\hline
\multicolumn{1}{r}{[Z/H]}  &   $-$1.63 &   $-$1.23 &   $-$0.63 &   $-$0.33 &    0.05 &    0.25 \\
log(age) \\
\hline
 8.00 &     0 &      6 &      0 &      0 &      0 &      0 \\
 8.25 &     6 &      0 &      0 &      5 &      0 &      0 \\
 8.50 &     0 &      0 &      0 &      0 &      0 &      0 \\
 8.75 &     0 &      0 &      0 &      0 &      0 &      0 \\
 9.00 &     0 &      0 &      0 &      0 &      0 &      0 \\
 9.25 &     0 &      0 &      0 &      1 &      0 &      0 \\
 9.50 &     0 &      0 &      0 &      0 &      0 &      4 \\
 9.75 &     0 &      0 &      0 &      0 &     36 &      0 \\
10.00 &     0 &      0 &      0 &      0 &      0 &      0 \\
10.18 &     0 &     15 &      0 &      0 &     27 &      0 \\
\hline
\end{tabular}
\end{center}
\end{table}

\begin{table}
\caption{Same as in \ref{tab_bc03_center}, but for position 2 and Vazdekis et al. models.}
\begin{center}
\begin{tabular}{lcccccc}
\hline
\multicolumn{1}{r}{[Z/H]}  &   $-$1.63 &   $-$1.23 &   $-$0.63 &   $-$0.33 &    0.05 &    0.25 \\
log(age) \\
\hline
 8.00 &     0 &      5 &      0 &      0 &      0 &      0  \\
 8.25 &    14 &      0 &      0 &      1 &      0 &      0  \\
 8.50 &     0 &      0 &      0 &      0 &      0 &      0  \\
 8.75 &     0 &      0 &      0 &      0 &      0 &      0  \\
 9.00 &     0 &      0 &      0 &      0 &      0 &      0  \\
 9.25 &     0 &      0 &      0 &      0 &      0 &      0  \\
 9.50 &     0 &      0 &      0 &      0 &      0 &      0  \\
 9.75 &     0 &      0 &      0 &      4 &      5 &      0  \\
10.00 &     0 &      0 &      0 &      0 &      0 &      0  \\
10.18 &     0 &     25 &      0 &      0 &     47 &      0  \\
\hline
\end{tabular}
\end{center}
\end{table}
 
\end{document}